\documentclass[onecolumn,authoryear]{els-mrw} 

\usepackage{amsmath,amssymb,amsfonts,amsthm,makeidx,graphicx}
\usepackage[varg]{txfonts}
\usepackage{helvet}

\usepackage{bm}

\def\aap{A\&A}%
\def\apj{ApJ}%
\def\apjl{ApJ}%
\def\mnras{MNRAS}%
\def\araa{Annual Rev. Astron. and Astroph}%
\def\icarus{Icarus}%
\def\planss{Planetary and Space Science}

\usepackage[version=4]{mhchem}
\usepackage{color}
\usepackage{xspace}
\newcommand{\eq}[1]{Eq.~(\ref{eq:#1})}
\newcommand{\eqp}[1]{\mbox{(Eq.\ \ref{eq:#1})}}
\newcommand{\Eq}[1]{\mbox{Equation~(\ref{eq:#1})}}

\newcommand{\eqsp}[2]{Eqs.\ \ref{eq:#1} and \ref{eq:#2}}

\newcommand{\se}[1]{\mbox{Sect.~\ref{sec:#1}}}
\newcommand{\Fg}[1]{\mbox{Figure~\ref{fig:#1}}}
\newcommand{\fg}[1]{\mbox{Fig.~\ref{fig:#1}}}

\newcommand{\tb}[1]{\mbox{Table~\ref{tab:#1}}}

\usepackage[colorlinks=true,linkcolor=blue,urlcolor=blue]{hyperref}

\begin{document}

\chapter{Planet Formation Mechanisms}\label{chap1}

\author[1]{Chris Ormel}%

\address[1]{\orgname{Tsinghua University}, \orgdiv{Department of Astronomy}, \orgaddress{
        30 Shuangqing Rd, Beijing 100084, China}}

\articletag{Chapter Article tagline: update of previous edition,, reprint..}

\maketitle

\begin{glossary}[Learning Objectives]
    \begin{itemize}
        \item Distinguish between the advantages and drawbacks of the core accretion and disk instability models of planet formation
        \item Understand the key physical principles that govern the initial, non-gravitational growth of particles
        \item Identify the major challenges associated with the formation of planetesimals
        \item Describe the characteristics of planetesimal accretion and pebble accretion
        \item Understand the challenges involved in forming planets at large distances from their host star
        \item Explain when and how atmospheres start to form around planetary bodies, and identify the factors limiting gas accretion
    \end{itemize}
\end{glossary}

\begin{glossary}[Glossary]
    \term{Accretion} The process by which matter accumulates onto a more massive body, usually through gravitational forces. In planet formation this entails the movement of gas onto the star or planet, as well as the growth of planetary bodies by accumulation of smaller solids.  \\
    \term{Aerodynamical size ($\tau_s$)} A dimensionless parameter indicating the level particles couple to the gas, with $\tau_s\ll1$ indicating tight coupling and $\tau_s\gg1$ negligible coupling.  \\
    \term{Cold start} A formation process of giant planets where planets start with low entropies, usually associated with the core accretion mechanism. \\
    \term{Collisional cascade} A process where large bodies undergo fragmenting collisions, progressively breaking down into smaller particles.\\
    \term{Core accretion} A model for giant planets in which a protoplanet forms from the solid material through a bottom-up process, before accreting gas from the protoplanetary disk. \\
    \term{Disk instability} A model where giant planets form directly from the collapse of gravitationally unstable regions within the disk. \\
    \term{Hill radius}  The approximate region around a planetary body within which its gravitational influence dominates over that of the star. The Hill radius extends through the L1 and L2 Lagrange points. \\
    \term{Hot start}  A formation process of giant planets where planets start with high entropies, usually associated with the disk instability mechanism. \\
    \term{Isolation mass} The maximum planet mass at which further growth ceases for a certain process.\\
    \term{Kelvin-Helmholtz instability} A fluid instability caused by a large velocity gradient in a direction perpendicular to the gas flow.\\
    \term{Kelvin-Helmholtz timescale} The time it takes for a self-gravitating body to cool and contract by radiation. \\
    \term{Meter size problem} The rapid radial drift of $\tau_s\sim1$ particles, leading to their loss to the star in about 100 orbital periods. \\
    \term{Oligarchic growth phase} The phase following runaway growth during which protoplanets accrete planetesimals at a slower pace, due to strong dynamical feedback (viscous stirring). \\
    \term{Pebbles} Aerodynamically active (loosely coupled) particles that can traverse large distances during the lifetime of the gas-rich protoplanetary disk.  \\
    \term{Planetesimals} Gravitationally bound bodies of size typically in the 1--100 km range. \\
    \term{Roche density} The density at which the self-gravity of a collection of particles can withstand tidal disruption from another massive body. \\
    \term{Runaway growth phase} A growth phase where dynamical feedback on the planetesimal population is limited, allowing protoplanets to grow at least exponentially with time. \\
    \term{Stopping time} The time gas friction needs to align the particle motion to that of the gas \\
    \term{Tidal downsizing} A process through which terrestrial planets form from DI-induced clumps. \\
    \term{Viscous stirring} The dynamical heating of bodies due to collisionless encounters.\\
\end{glossary}

\begin{glossary}[Nomenclature]
\begin{tabular}{@{}lp{34pc}@{}}
CA      &Core accretion\\
DI      &Disk instability\\
H+He    &Hydrogen and helium gas\\
ISM     &Interstellar medium\\
SGI     &Secular gravitational instability\\
MMSN    &Minimum-mass solar nebula \\
\end{tabular}
\end{glossary}

\begin{abstract}[Abstract]
    Planet formation encompasses processes that span a remarkable 40 magnitudes in mass, ranging from collisions between micron-sized grains inherited from the ISM to the accretion of gas by giant planets. The planet formation process takes place in the interior of dusty disks, which offer us only limited observational constraints. Historically, the two main paradigms describing planet formation are the disk instability and core accretion models. In the former giant planets condense directly form a disk massive enough to fragments under its own self-gravity. In contrast, the core accretion model follows a bottom-up approach driven by the growth of solid bodies. The core accretion model includes the direct growth phase, where pebble and possibly planetesimal sized bodies form by surface-driven processes. Alternatively, planetesimal bodies can emerge from instabilities in the solid population, leading to the formation of clumps that ultimately collapse. The later stages of the core accretion model involve the emergence of protoplanets from the planetesimal population, which accretes smaller bodies|planetesimals or pebbles|before ultimately accreting freely the gas of the disk. In this Chapter I will review these formation mechanisms, aiming, where possible, to build an intuitive understanding from elementary physical principles.
\end{abstract}

\section{Introduction}
Planet formation is a long-standing problem in astrophysics. One of the earliest descriptions of the process is the ``nebular hypothesis'', proposed by Kant and Laplace. While developed inthe 18$^\textrm{th}$ century, its core principles are broadly followed today. In this hypothesis the star and planets form concurrently, with planets emerging from a disk that surrounded the newly formed star.  A key challenge for planet formation theory is the difficulty to test predictions with direct observational evidence. This is because: (i) planets form in the midplane regions of a dust-enshrouded disk; (ii) the small angular scales are difficult to resolve and the process is, by astrophysical standards, not particularly energetic; and (iii) the timescales are relatively short|a mere $\sim$Myr, with some key processes potentially happening even faster. For these reasons, observational constraints tend to be indirect. Young forming planets can alter the properties of the surrounding disk, though direct detection of forming planets is hard. Additionally, the physical and dynamical properties of exoplanet systems may carry imprints from their formation era, and the fossils from the solar system in the form of (undifferentiated) meteorites provides further insights.

To keep this chapter focused, I will limit the discussion to planet formation \textit{mechanisms} and avoid covering observational aspects.  Further, I will largely omit other important topics as planet migration, planet population synthesis, planet dynamics, and disk-related processes. For a more comprehensive overview of recent progress, I refer the reader to review chapters, e.g., those compiled in \textit{Protostars and Planets VII}\footnote{\url{http://ppvii.org/chapter-list/index.html}}.  Even so, discussion of the formation mechanisms in this chapter is cursory; we will only scratch the surface of this complex topic.

\begin{figure}[t]
\centering
\includegraphics[width=\textwidth]{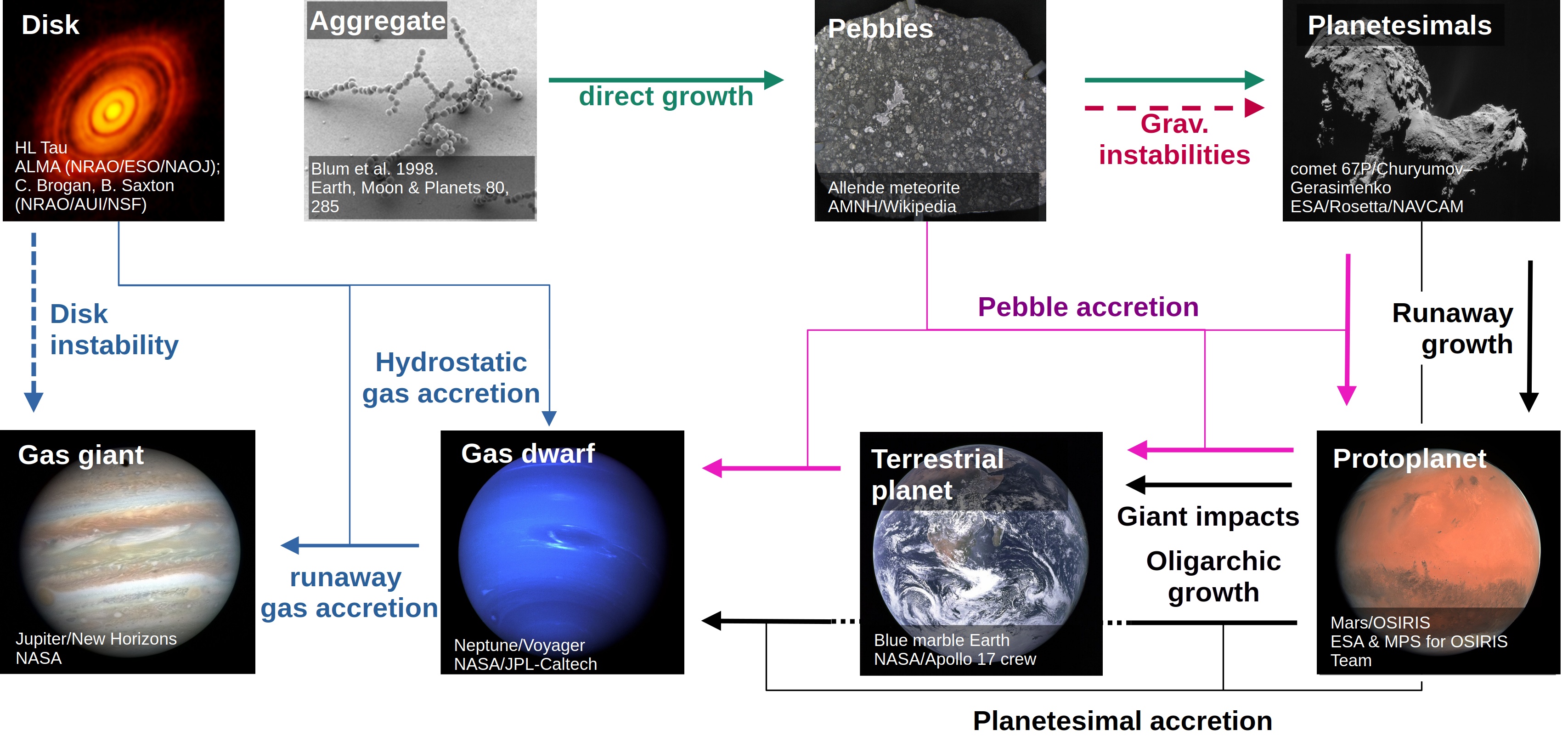}
\caption{Routes to form planetary bodies. Mechanisms are indicated with an arrow and labeled. Gravitational instabilities (involving a collection of particles or gas) are indicated by dashed arrows. Accretion processes that rely on surface forces are in green, gravitational forces in black, gas accretion in blue, and pebble accretion (which involves both gas and gravity) is in magenta. The disk instability (DI) mechanism is depicted at the very left. Other processes fall under the core accretion (CA) umbrella.}
\label{fig:taxonomy}
\end{figure}
\Fg{taxonomy} presents an overview of the various formation processes. The leftmost arrow depicts planet formation through disk instability (DI), where gas giant planets form directly from clumps in a gravitationally unstable disk (\se{DI}). In contrast, all other processes are part of the \textit{core accretion} (CA) model, where a ${\sim}1$--$10$ Earth-mass ``core'' forms before gas is accreted. One advantage of the CA model is that it explains the emergence of a wide range of planetary objects, not just gas giants. Core accretion (CA), however, is an umbrella term that encompasses a plethora of physical processes, spanning roughly 40 orders of magnitude in mass, from sub-micron sized grains to Earth-sized planets. First, grains coagulate into dust aggregates and pebbles, driven by weak surface forces (\se{direct}). Sticking between particles require the dissipation of kinetic energy during the collision, which becomes harder with increasing mass and velocity. Growth likely ceases at sizes in the ${\sim}100\,\mu\mathrm{m}$-to-$\mathrm{cm}$ range, consistent with the inferred particle sizes of the continuum emission from many disks. Another obstacle to growth is the so-called meter-size barrier (see \se{prelim}) where bodies (not necessarily of meter size) drift to the star on ${\sim}100$ dynamical timescales. Some models argue that the ``hit-and-stick'' growth can lead to planetesimals, but the more commonly invoked mechanism for the emergence of planetesimal is the concentration and subsequent collapse of pebbles (\se{instability}).

For planetesimals and larger bodies, self-gravity becomes the dominant binding mechanism. Gravity is also instrumental in speeding up growth by significantly increasing the two-body cross section for collisions (\se{pltsm-planets}). In the classical model, developed primarily with the solar system in mind, a series of protoplanets form out of the planetesimal population through runaway growth, followed by the slower oligarchic accretion mode. More recently, it has been realized that smaller pebble-sized particles can contribute substantially to growth. In reality, planetesimal and pebble accretion likely operate in tandem. Once these smaller building blocks are gone and the disk has dissipated, protoplanets may merge through a series of giant impacts to ensure the long-term stability of the planetary system. For the Earth, the Moon-forming impact is considered the last in this sequence.

Large enough protoplanets can also bind the primordial (i.e., H+He-dominated) gas of the disk. Initially, this proceeds under conditions of hydrostatic balance, where the protoplanet envelope is in pressure equilibrium with the disk gas. Over time this envelope will grow either because the mass of the planet's metallic component increases or because the planet envelope cools, eventually reaching the point where hydrostatic balance can no longer be maintained. If the primordial disk is still present at this stage, H+He gas will be accreted violently or be funneled through a circumplanetary disk. The final mass of the planet is determined by the ability of the planet to carve a gap in the disk, as well as the total amount of gas in its vicinity and the disk lifetime. 

\subsection{Preliminaries}
\label{sec:prelim}
Although disk processes are not within the scope of this chapter, a basic description of disks is essential as planet formation mechanisms rely on the state of the gas. Historically, disks have been modelled as power-laws
\begin{equation}
    \Sigma_\mathrm{gas} = \Sigma_0 \left( \frac{r}{\mathrm{au}} \right)^p
    \label{eq:sigma-gas}
\end{equation}
where $p=-3/2$ and $\Sigma_0=1\,700\,\mathrm{g\,cm^{-2}}$ are adopted in case of the minimum-mass solar nebular model or MMSN \citep{Hayashi1981}. The MMSN profile for the gas surface density $\Sigma_\mathrm{gas}$ is obtained from the current distribution of metals contained in the planets of the solar system and scaling it by a solids-to-gas ratio. While this approach is somewhat dated---it does not account for effects as particle drift or planet migration---MMSN-like disk models still serve as useful benchmarks. Steady-state viscous accretion results in $p=-1$. When accretion is driven by disks winds, other prescriptions may apply. The vertical dimension of a disk is often modeled as isothermal. This approximation certainly fails in the upper disk regions, but may be appropriate for the disk midplane, where planets form. The isothermal assumption, combined with hydrostatic balance, result in a normal distribution for the gas density:
\begin{equation}
    \rho_\mathrm{gas}(z) = \frac{\Sigma_\mathrm{gas}}{H\sqrt{2\pi}} \exp \left[ -\frac{1}{2} \left( \frac{z}{H} \right)^2 \right]
\end{equation}
where $z$ is the distance from the midplane, $H = \left.\sqrt{k_B T /m_\mathrm{gas}}\right/\Omega_K$ is the pressure scaleheight with $k_B$ Boltzmann's constant, $T$ temperature and $m_\mathrm{gas}$ the mean mass of the gas molecules. In many disk models disks are flared: the disk aspect ratio, $h=H/r$, increases with $r$.

In the radial direction, the stellar gravity is mostly supported by rotation, but pressure support plays a crucial role for particle aerodynamics. Specifically, let
\begin{equation}
    \eta \equiv -\frac{1}{2\rho_\mathrm{gas} \Omega_K^2 r}  \frac{\partial P}{\partial r} \sim h^2 
    \rightarrow 9\times10^{-4}
    \label{eq:eta}
\end{equation}
i.e., $\eta$ is half the ratio of the hydrostatic force over the stellar gravity.  In reducing the expression I approximated $\partial_r P \sim -P/r$ and inserted the ideal gas law.  Finally, the $\rightarrow$ indicates that the expression is evaluated by inserting the numerical values compiled in \tb{params} (in this case only $h$). These values are appropriate for 1\,au. For flared disks, the value of $\eta$ will therefore increase in the outer disk.  

Hence, the azimuthal velocity $v_{\phi,\mathrm{gas}} \simeq (1-\eta)v_K$. Due to the slight pressure support, the gas rotates sub-Keplerian by an amount $\eta v_K \sim 30\,\mathrm{m\,s^{-1}}$. When the pressure gradient (locally) reverses ($\eta<0$), rotation will instead be super-Keplerian.  Particles in disks interact aerodynamically or gravitationally with the gas. The conditions in the protoplanetary disk vary such that the gas drag force takes on a variety of forms, ranging from Newtonian (quadratic in velocity) at large sizes to Stokes (linear in velocity) to Epstein (free molecular flow; also linear in $\Delta v$) at small sizes. It is more common to express the aerodynamical interaction in terms of a \textit{stopping time} $t_\mathrm{stop}=m\Delta v/F_D$, where $m$ is the particle mass, $\Delta v$ the relative particle-gas velocity, and $F_D$ the gas drag force. In the case of a drag law linear in velocity $t_\mathrm{stop}$ is independent of velocity and can be identified with a particle size. In this regard, the dimensionless stopping time $\tau_s=\Omega_K t_\mathrm{stop}$, sometimes referred to as Stokes number or the aerodynamical size, is particular relevant. 

In disks, particles do not feel the gas pressure ($\rho_\mathrm{gas}/\rho_\bullet \ll 1$) and tend to orbit faster than the gas. Hence, they experience a headwind, which extracts angular momentum from particles, resulting them to drift towards the star. Solving for the Euler equations of particles and gas simultaneously, the expressions for the particle radial drift and azimuthal motion follow \citep{NakagawaEtal1986}: 
\begin{equation}
    \label{eq:drift-motions}
    v_r = -\frac{2\tau_s}{(1+Z)^2 +\tau_s^2} \eta v_K; \qquad
    v_\phi -v_K = -\frac{1+Z}{(1+Z)^2 +\tau_s^2} \eta v_K; \qquad
    v_\mathrm{\phi,gas} -v_K = -\frac{1+Z+\tau_s^2}{(1+Z)^2 +\tau_s^2} \eta v_K
\end{equation}
where $Z$ is the midplane solids-to-gas ratio and $v_\mathrm{\phi,gas}$ is the azimuthal velocity of the gas. If $Z\gg1$ or $\tau_s\gg1$ these velocities go to zero as particles do not experience much gas drag. Particles of $\tau_s=1$ are the fastest drifters and get removed in ${\sim}100$ local orbital periods. This problem where particles get flushed into the star by aerodynamical forces in smooth disks has been known as the meter size problem.  Conversely, these mobile particles can also assist in triggering planetesimal formation and aiding the growth of planetary bodies in the inner disk regions.  In the following, we refer to these aerodynamically active particles|particles that can drift over significant distances within the lifetime of the disk (${\sim}10^6$--$10^7$ yr)|as pebbles.

\begin{table}[t]
    \TBL{\caption{\label{tab:params}Variables and parameters used throughout the Chapter.}}
    {\begin{tabular*}{\textwidth}{@{}llp{5.5cm}lp{5.5cm}}
\toprule
\multicolumn{1}{@{}l}{\TCH{Parameter}} &
\multicolumn{1}{l}{\TCH{Value}} &
\multicolumn{1}{l}{\TCH{Description}} &
\multicolumn{1}{@{}l}{\TCH{Parameter}} &
\multicolumn{1}{l}{\TCH{Description}}\\
\colrule
%$\rho_\bullet$                      & $5\,\mathrm{g\,cm^{-3}}$            & | (planet)         \\
%$\Sigma_\mathrm{gas}r^2/m_\star$    & $0.1$            & | (planet)         \\
$\Delta v$                          & $30\,\mathrm{m\,s^{-1}}$            & planet-pebble relative velocity                             & $\Delta v$    & relative velocity \\
$\Sigma$                            & $10\,\mathrm{g\,cm^{-2}}$           & surface density in solids                                   & $\Theta$      & gravitational focusing factor\\
$\Sigma_\mathrm{gas}$               & $10^3\,\mathrm{g\,cm^{-2}}$         & surface density in gas                                      & $\Omega_K$    & Keplerian orbital frequency \\
$\gamma$                            & $10\,\mathrm{erg\,cm^{-2}}$         & surface energy density                                      & \\
$\delta$                            & $10^{-4}$                           & particle diffusivity parameter $(=D/H^2\Omega)$             & $\eta$        & dimensionless form radial pressure gradient \\
$\kappa$                            & $1\,\mathrm{cm^2\,g^{-1}}$          & opacity  (\se{gas})                                         & $\kappa$      & epicycle frequency (\se{DI}) \\
$\rho_\bullet$                      & $1\,\mathrm{g\,cm^{-3}}$            & internal density                                            & $\sigma$      & collisional cross section (\se{instability}: surface density) \\
$\rho_{\bullet,p}$                  & $5\,\mathrm{g\,cm^{-3}}$            & planet internal density                                     & $\sigma_{90}$ & cross section for 90-degrees scattering \\
$\tau_s$                            & $0.1$                               & pebble aerodynamical size                                   & $G$           & Newton's gravitational constant \\
$R_g$                               & $1\,\mu\mathrm{m}$                  & grain radius                                                & $H$           & disk scaleheight (gas) \\
$R_s$                               & $1\,\mathrm{km}$                    & planetesimal radius                                         & $Q_T$         & Toomre-Q \\       
$R_p$                               & $10^4\,\mathrm{km}$                 & planet radius                                               & $R_B$         & Bondi radius \\
$T_\mathrm{disk}$                   & $100\,\mathrm{K}$                   & disk temperature                                            & $R_H$         & Hill radius \\ 
$\tilde{b}$                         & $10$                                & separation planetary bodies in terms of Hill radii          & $Z$           & midplane solids-to-gas ratio \\
$h$                                 & $0.03$                              & disk aspect ratio ($=H/r$)                                  & $e$           & eccentricity \\
$r$                                 & $1\,\mathrm{au}$                    & disk orbital radius                                         & $i$           & inclination \\
$\overline{m}_\mathrm{gas}$         & $2.34\,m_\mathrm{u}$                & mean molecular mass                                         & $n$           & number density \\
$m_p$                               & $1\,m_\oplus$                       & planet mass                                                 & $v_K$         & Keplerian velocity    \\
$m_\star$                           & $1\,m_\odot$                        & stellar mass                                                & $v_\phi$      & azimuthal velocity    \\
$\mathcal{E}$                       & $10^{10}\,\mathrm{erg\,cm^{-3}}$    & Young's modulus                                             & $v_r$         & radial (drift) velocity \\
\botrule
\end{tabular*}}{%
\begin{tablenotes}
    \footnotetext{
        In the left column, representative numerical values are provided. Values correspond approximately to conditions of a quiescent disk at 1\,au for a class II, solar-type star. Whenever $\rightarrow$ appears in the main text these values are used to evaluate the preceeding expression.}
\end{tablenotes}
}%
\end{table}

\section{Disk Instability}
\label{sec:DI}

Inviscid disks are governed by three key equations: the continuity equation (mass conservation), the Euler equation, which describes force balance, and the equation of state, which describes a relationship between pressure, density and temperature. Ignoring the disk self-gravity, a solution that satisfies these equations can always be found. However, the disk cannot be arbitrarily massive; when the gas surface density exceeds a threshold, local patches collapse in much the same way as clouds fragments.  This process can be described by a linear stability analysis, where a relation between the scale of the perturbation $\lambda$ and the growth rate $\omega$ is sought for. For axisymmetric perturbations in the thin disk approximation (the disk scaleheight exceeds $\lambda$) the dispersion relationship reads \citep[e.g.,][]{BinneyTremaine2008}
\begin{equation}
    \omega^2 = \kappa^2 -2\pi G \Sigma_\mathrm{gas} k + c_s^2 k^2
    \label{eq:dispersion}
\end{equation}
where $k=2\pi/\lambda$ and $\kappa$ is the epicycle frequency, which equals the orbital frequency $\Omega_K$ for Keplerian disks, $c_s$ is the sound speed of the gas, $G$ Newton's gravitational constant, and $\Sigma_\mathrm{gas}$ the surface density. Disk Instability (DI) ensues when $\omega^2<0$ (conceptually: the destabilizing gravity term on the RHS outweighs the stabilizing contributions from rotation and pressure). 
This can be expressed by the Toomre-Q criterion:
\begin{equation}
    \label{eq:Toomre-Q}
    Q_T \equiv \frac{c_s\Omega_K}{\pi G\Sigma_\mathrm{gas}}
    = \frac{h m_\star}{\pi \Sigma_\mathrm{gas} r^2}
    \rightarrow 85
\end{equation}
where we have used $c_s=H\Omega_K$. The threshold for DI is $Q_T \lesssim 1$, which implies that for disks to fragment they need to be cool (small $h$) and massive (high $\Sigma_\mathrm{gas}$). Wherever $Q_T\lesssim1$ patches of scale $\lambda_c\sim c_s^2/G\Sigma_\mathrm{gas}$ become unstable, corresponding to a mass of
\begin{equation}
    m_c \sim \lambda_c^2 \Sigma_c 
    \sim Q_T  h^3 m_\star
    \rightarrow
    0.03 Q_T\,m_\mathrm{J}
%   = 1\,m_\mathrm{J}\; Q_T^4 \left( \frac{\Sigma_\mathrm{gas}r^2/m_\star}{0.1} \right)^3.
\end{equation}
It should be emphasized that DI ($Q_T\lesssim1$) does not guarantee clumps collapse into Jupiter-mass planets. 
To form planets, clumps need to cool and contract; collapse will be resisted if the energy released by the collapse cannot be radiated away. In that case the clump will simply be sheared out. In other words, for a sustained collapse the disk cooling time $t_\mathrm{cool}$ must be short enough compared to the orbital timescale. Typical values, expressed in terms of a $\beta$ parameters, point to $\beta = t_\mathrm{cool}\Omega_K \lesssim 10$ \citep{KratterLodato2016}.
Furthermore, even if the cooling condition is met, gravitational interactions with e.g., other clumps or a spiral density wave may result in their dispersal. Another question is whether the clumps, if they survive, would stay at planetary masses ${\sim}m_J$ or whether they continue to accrete gas to end up as Brown dwarfs. To arrest accretion, planets must open a gap, which is harder to achieve in a disk which is by definition ``active'' due to it being gravitationally unstable.  

Planets formed by DI become giants. To provide a channel for the formation of terrestrial planets, the clumps need to migrate inwards where their Hill radius
\begin{equation}
    R_\mathrm{H} \equiv r \left( \frac{m_p}{3m_\star} \right)^{1/3}
    \rightarrow 0.01\,\mathrm{au}
\end{equation}
is much smaller, such that it can sheds its outer gaseous layers. This is the tidal downsizing scenario \citep{Nayakshin2010}. For it to work clumps should survive the inward migration, not contract too much, while the solids inside the clump must efficiently sediment to its center.

Whether or not DI is a viable planet formation scenario depends on its ability to explain trends in the exoplanet census (not the focus of this review). One long-standing trend is the correlation of giant planet occurrence with stellar metallicity, which suits the core accretion scenario more naturally. Of course, both scenarios could be operating with DI the preferred mechanism to form planets at distance ${\gtrsim}10\,\mathrm{au}$ where the core accretion process may take too long (see however \se{pebacc}). Another way to distinguish planets formed by DI from CA is in the initial entropy that these planets start off with.  During runaway gas accretion, gas shocks onto the surface of a giant planet, and the energy from the shock-heated gas is rapidly radiated away. As a result, giant planets formed by CA end up with low entropy after formation, referred to as a cold start. On the other hand, planets formed by DI, where the collapse is by slow Kelvin-Helmholtz contraction, retain more heat, resulting in a hot start. During their early evolution hot-start planets would appear to be brighter.

%which is not a prime\? outcome of the DI mechanism. It could be a secondary feature\? if processes like cooling and settling depend on disk metallicity, 

\section{Direct growth}
\label{sec:direct}
\begin{figure}[t]
\centering
\includegraphics[width=.9\textwidth]{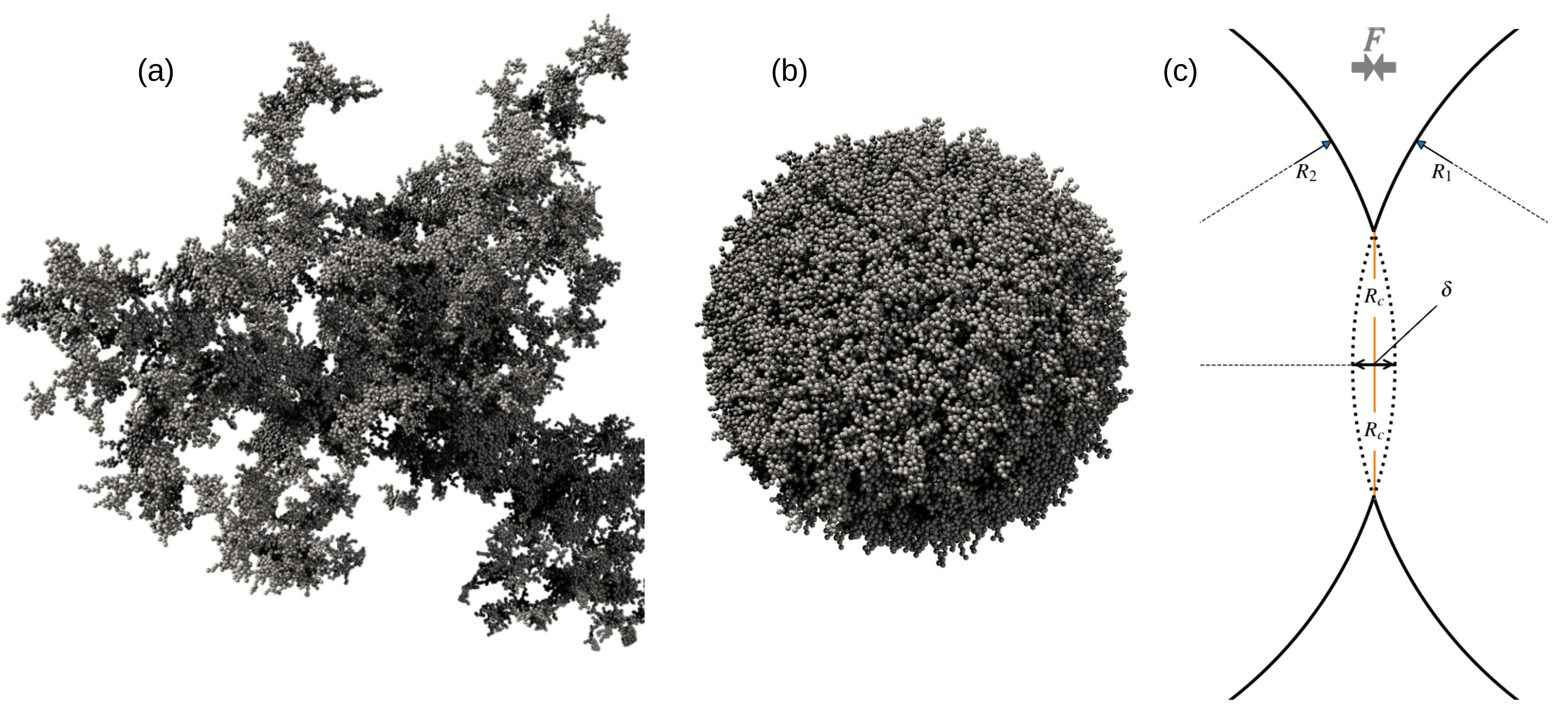}
\caption{\label{fig:aggregates}(a+b) Examples of fractal and homogeneous (compacted) aggregates \citep{SeizingerEtal2013}. Fractal growth describes a growth process in which mass and radius are related as $m\propto R^{D_f}$ where $D_f$ is the fractal dimension. (c) Geometry of the contact area. The attractive force is provided by the surface interaction.}
\end{figure}
With direct growth we mean the gradual increase of the mass of a particle by surface-related processes. This could be the deposition of vapor on the surface of particles or the sticking of two solid particles through surface forces upon contact. For equal-mass particles this collisional aggregation would double the mass of the particles. The efficacy of direct growth are determined by the particle sizes, the collisional energy, and the material properties of the surface area. It is generally accepted that direct growth can produce at least pebble-sized particles, but to what size the realm of direct growth extends is a key, open question.

\subsection{Particle growth by condensation}
Growth by condensation, or more precisely deposition (the direct conversion of vapor to the solid phase), occurs when the vapor is saturated, meaning that the partial pressure of a species $Z$ exceeds its saturation vapor pressure $P_{\mathrm{sat},Z}(T)$. Condensation is particularly relevant in the cool, outer disk regions, where volatiles (e.g, \ce{CH4}, \ce{CO}, and \ce{H2O}) as well as nobel gases can condense onto grains. However, it is unlikely that \ce{H2} will condense.  Condensation forms ice layers on refractory grains, which may increase the mass of the solid component and alter the sticking properties. Inside condensation fronts (icelines), the frozen-out volatiles can be returned to the gas phase.

\subsection{Physics of collisional aggregation}
A more potent growth mechanism are binary collisions.  
%, because this is energetically favored -- it takes energy to create surface. Because the amount of energy that can be dissipated is proportional to the surface area, whereas the collisional energy is proportional to mass, the efficacy of direct growth is expected to decrease with size. 
When two spherical particles are brought into contact, strain-induced stresses create a repulsive force, determined  by Young's modulus $\mathcal{E}$, which quantifies the elasticity of the material (materials with lower $\mathcal{E}$ are more elastic).
%to form a common contact area, this strain-induced stresses would repel the particles with a force determined. 
However, forming a contact area is energetically favorable as it reduces the overall surface energy. \citet{JohnsonEtal1971} showed that the equilibrium contact area is $A_\mathrm{eq}\sim (\gamma R_g^2/\mathcal{E})^{2/3}$ where $R_g$ is the reduced grain size and $\gamma$ the surface energy density (see \fg{aggregates}c). Therefore, an energy $E_\mathrm{br} = \gamma A_\mathrm{eq}$ is required to break this contact. The formation of a contact is facilitated by the asymmetric nature of the interaction. During ingress a contact forms when the distance between the two particles equals the sum of their radii, $d=d_\mathrm{in}=R_\mathrm{g1}+R_\mathrm{g2}$. During egress, however, the contact persists for $d>d_\mathrm{in}$, connecting the two grains by a material neck that tends to pull them back. Equating the breaking energy $E_\mathrm{br}$ with the collision energy for equal-sized particles, we obtain a sticking threshold of:
\begin{equation}
    \label{eq:vstick}
    v_\mathrm{stick} \sim \left( \frac{\gamma^{5}}{\mathcal{E}^{2} R_g^{5} \rho_\bullet^{3}} \right)^{1/6}
    \rightarrow 7\,\mathrm{cm\,s^{-1}}.
\end{equation}
Softer materials (lower $\mathcal{E}$) and those with higher $\gamma$ promote sticking. Particles that form through this ``hit-and-stick'' mechanism form aggregates, possibly of low fractal dimension, characterized by large voids (e.g., \fg{aggregates}a). 

Apart from breaking contacts, the collisional energy can also be directed to restructure aggregates. Consider the translational motion of two particles over a contact (rolling). It can be shown that the contact lags the rolling motion by a distance $\xi_\mathrm{crit}$ (estimated to be ${\sim}1$--10\AA), which results in a resistive force opposing the motion. The energy required to move the contact over a 90$^\circ$ angle is known as the rolling energy, $E_\mathrm{roll}\sim \gamma \xi_\mathrm{crit} R_g$ \citep{DominikTielens1997}. Restructuring of aggregates occurs when the collisional energy exceeds $E_\mathrm{roll}$, causing aggregates to compact (e.g., \fg{aggregates}b). In addition, aggregates are expected to be disrupted when the collisional energy surpasses ${\sim}n_c E_\mathrm{br}$ where $n_c$ is the number of contacts in an aggregate.

Finally, bouncing is another collisional outcome. Laboratory experiments in particular have demonstrated that homogeneous, macroscopic particles are prone to bounce at the collisional velocities typical of the protoplanetary disk environment \citep{GuettlerEtal2010}. 

\subsection{The outcome of direct growth}
It is likely that coagulation first operates in the ``hit-and-stick'' mode, characterized by the formation of large fractal aggregates. The emergence of this fractal structure results in particles with a large geometrical cross section compared to their mass. The high surface area-to-mass ratio renders the aggregates tightly coupled to the gas, which suppresses their relative velocity. However, once the collisional energy $E_\mathrm{coll}$ exceeds the rolling energy $E_\mathrm{roll}$ particles starts to compact. The effectiveness of the collisional compaction is, however, a matter of debate. Two pathways can be envisioned.

In the first scenario, collisional compaction is self-reinforcing. Any compaction event would increase the particle's stopping time, which would in turn increase $\Delta v$, promoting ever-more energetic collisions. This pathway has been developed based on theoretical considerations as well as laboratory experiments. The end product of this growth route would be a distribution where most of the mass is in relatively compact, pebble size particles, with no further growth due to the dominance of bouncing \citep{ZsomEtal2010}. 

Conversely, it has been argued, based on a series of numerical experiments, that collisional compaction of fractal aggregates is in fact weak. That is, the creation of new voids due to collisions more than offsets the voids lost by compaction. In fact, the compacted material still follows a fractal structure. Consequently, in this scenario aggregates stay highly porous, allowing growth to outpace the meter-size barrier and form planetesimals very rapidly \citep{OkuzumiEtal2012}. When these fluff balls become massive enough they will ultimately compact through gas drag and self-gravity, reaching densities similar to those of comets \citep{KataokaEtal2013i}.

Several factors may complicate our understanding on the outcome of direct growth. First, collisional outcomes are highly sensitive to material parameters, such as $\gamma$, $\mathcal{E}$, and $R_g$. There are indications that these are in fact not constants but depend on, e.g., temperature \citep[e.g.][]{MusiolikWurm2019}. Collisions between particles of very different size can also complicate the picture. It has been argued that for compact particles this is a possible channel to overcome the meter-size barrier \citep{WindmarkEtal2012}, whereas for fractal particles erosive collisions may inhibit growth \citep{KrijtEtal2016}. In summary, our understanding of the outcome of this crucial growth phase remains poor.

\section{Planetesimal formation by gravitational instability}
\label{sec:instability}
If direct growth of solids stalls, larger bodies (planetesimals) could alternatively form by gravitational instability of solids. We can distinguish three stages:  
\begin{itemize}
    \item \textbf{Concentration}. Disk-related processes, in conjunction with particles' aerodynamical properties, cause solids to converge in either the vertical, azimuthal, or radial dimension, raising the local solids-to-gas ratio.
    \item \textbf{Instability}. Possibly aided by an elevated solids-to-gas ratio, an instability in the particle layer could be triggered, which could further elevate solid densities. Generally, instabilities proceed in a linear phase where densities grow proportional to time and a nonlinear phase, in which the instability may saturate (i.e., reach peak densities) or continue to grow. 
    \item \textbf{Collapse}. This addresses the question whether the instability produces gravitationally bound clumps that collapse to solid densities or whether the clumps disperse before they have the opportunity to collapse.
\end{itemize}
There are a great number of mechanisms to concentrate particles, and perhaps even more instabilities. Producing a full, comprehensive review is beyond the scope of this chapter and what follows is a selective overview that broadly follows the above sequence.

\subsection{Particle concentration}
Loosely coupled particles drift in the direction of the excess gravity at velocities $\bm{v}=\bm{g}t_\mathrm{stop}$. Any disk instability that creates pressure gradients will therefore have a response in the solid population. For example, anticyclonic vortices are high pressure regions in which particles concentrate. Acting on smaller scales, turbulence has the opposite effect, concentrating particles of stopping time equal to the Kolmogorov time into regions of low vorticity. 

In addition to these local concentration mechanisms particles can also be concentrated due to drift motions.

\subsubsection{Vertical concentration}
The vertical component of the stellar gravitational force causes particles to settle on a time $t_\mathrm{settl}=z/v_\mathrm{settl}\sim \Omega_K^{-1}/\tau_s$ (for $\tau_s<1$), forming a solid-rich midplane. Two obstacles limit the amount of settling, however. First, any amount of turbulence in disks would oppose dust settling. The degree of settling can be obtained by equating the settling time with the diffusion time:
\begin{equation}
    H_\mathrm{solid} = \sqrt{\frac{\delta}{\delta +\tau_s}} H_\mathrm{gas}
    \rightarrow 0.03 H_\mathrm{gas}
    \label{eq:Hdust}
\end{equation}
where $\delta$ expresses the diffusivity: $D_z = \delta H^2/\Omega_K$. Second, even for globally quiescent disks, the solid-dominated midplane may trigger Kelvin-Helmholtz (KH) instabilities. Due to particle settling, the midplane solid-to-gas ratio $Z$  would be raised correspondingly, by a factor ${\sim}H/H_\mathrm{solid}$. When $Z\gg1$ the solid-dominated sub-disks start to rotate at Keplerian velocities, and it would drag the gas along \eqp{drift-motions}. However the gas disk above the midplane still rotates at the sub-Keplerian velocity, resulting in a vertical shear $\partial_z v_\phi$. Stability is measured in terms of the Richardson number 
\begin{equation}
    \mathrm{Ri} = \frac{g_z \partial_z \log\rho}{(\partial_z v_\phi)^2}
    \label{eq:Ri}
\end{equation}
which is a measure of stabilizing buoyancy over destabilizing shear. The layer will turn unstable when $\mathrm{Ri}$ falls below a critical value, $\mathrm{Ri}_\mathrm{crit}=1/4$, arresting the further settling of particles. It has been argued, however, that the density $\rho(z)$ will rearrange such that marginally stability $\mathrm{Ri}=\mathrm{Ri}_\mathrm{crit}$ is maintained. Under the single fluid assumption (and ignoring global turbulence, $\delta=0$), this adjustment leads to the formation of a density ``cusp'' near the midplane \citep{Sekiya1998}, which could trigger gravitational instability.

\subsubsection{Radial concentrations -- Aerodynamical pileup}
Another mechanism to concentrate particles is through convergent radial motions. 
%If particles sizes are fixed, $v_r$ \eqp{drift-motions} likely decreases with decreasing $r$. 
This can be understood by writing the continuity equation in Lagrangian form:
\begin{equation}
    \frac{D\Sigma}{Dt} = - \Sigma\frac{\partial}{\partial r} (r v_r).
    \label{eq:continuity}
\end{equation}
Hence, the surface density of a parcel increases with its inward drift when $-rv_r$ increases with $r$. If particle sizes are fixed then $t_\mathrm{stop}$ likely increases with decreasing gas density.  Particle convergence raises the solid density in the inner regions of the disk. However, if particle sizes are limited by drift or fragmentation processes, drift motions are rather determined by $\tau_s$ (i.e., particle sizes increase with decreasing radius). The evolution of solids in disks then proceeds inside-out with $\Sigma$ decreasing uniformly throughout. 

Disk processes can also effectively stall particles' inward drift. Any mechanisms that locally reverse the pressure gradient ($\eta<0$) would ``trap'' particles. Mechanisms that have been suggested to produce pressure maxima include gap opening by planets, zonal flows by magneto-rotational instability \citep{UribeEtal2011}, and shadows \citep{WuLithwick2021}. 

Evaporation fronts|icelines|are locations where drift motions could also produce pileups. The main reason is that the velocity at which particles drift towards the iceline much exceeds the gas velocity through which vapor is removed. This results in a pileup of vapor just interior to the iceline, part of which will be transported upstream across the iceline due to turbulent diffusion, where it can condense on the ice particles \citep{CuzziZahnle2004}. This accumulation renders the \ce{H2O} iceline an attractive location to form planetesimals. In addition, at temperatures close to its melting temperature, the wet nature of ice can enhance its ability to stick, promoting further aggregation. Interior to the snowline, there is another opportunity for pileups as the disaggregated particles have smaller $\tau_s$ and move at a lower radial velocity than their progenitor ice-rich pebbles \citep{IdaGuillot2016}. 

\subsection{Instability}
The dispersion relationship \eq{dispersion} also applies to solids. If particles do not couple with the gas, the sound speed can safely be put to zero in the dispersion relationship, resulting in all scales below $\lambda_c \sim G\Sigma/\Omega_K^2 \sim (r^2 \Sigma/m_\star) r$ becoming unstable. This is known as the Goldreich-Ward (GW) instability \citep{GoldreichWard1973}. The caveat is that particles must settle into a layer thinner than $\lambda_c$, which foremost requires a very quiescent disk. Using \eq{Hdust}:
\begin{equation}
    \delta < \left( \frac{r^2 \Sigma}{m_\star} \right)^2 \frac{\tau_s}{h^2}
    \rightarrow 1.4\times10^{-10}.
\end{equation}

The GW-mechanism ignores the effects of gas drag, which can both suppress and expedite instability.  Particles that are very tightly coupled to the gas form a single fluid with the gas (this requires $t_\mathrm{stop}$ to be less than any other relevant timescale).  A modified Toomre criterion for instability can be derived with the sound speed reduced according to the solids-to-gas ratio.  Under these conditions, the condition for instability much exceeds the Roche density \citep{ShiChiang2013}, see \eq{Roche}.

Relaxing the single fluid approximation is useful to consider the effect of the separate motion between gas and solids.  A slower axisymmetric instability is the secular gravitational instability (SGI). If there exists a surface density excess $\sigma$ (${\ll}\Sigma$) in a ring of radial scale $\lambda$, the gravity at the ring edges is $g\sim G\sigma$ and the settling velocity $\dot{\lambda}\sim -G\sigma t_\mathrm{stop}$. Using mass conservation ($\partial_t (\Sigma\lambda) = 0$) then gives the growth rate:
\begin{equation}
    \label{eq:sgi}
    t_\mathrm{sgi} = \frac{\sigma}{\dot{\sigma}} 
    \sim \frac{\lambda}{G\Sigma t_\mathrm{stop}}
    \sim \frac{m_\star}{\Sigma r^2}\frac{h}{\tau_s}\frac{\lambda}{H} \Omega_K^{-1};
    %\sim Q_T \frac{\Sigma_\mathrm{gas}}{\Sigma}\frac{\lambda}{H} \frac{\Omega^{-1}}{\tau_s};
    %\sim 4\times10^4\,\mathrm{yr} \times \frac{\lambda}{H}
    \qquad
    t_\mathrm{sgi,dif} \sim \delta \left( \frac{h}{\tau_s} \frac{m_\star}{\Sigma r^2} \right)^2 \Omega_K^{-1} 
    %t_\mathrm{sgi,dif} \sim \left( \frac{Q_T}{\tau_s} \frac{\Sigma_\mathrm{gas}}{\Sigma} \right)^2 \delta \Omega^{-1} 
    \rightarrow 1.1\times10^6\, \mathrm{yr}
\end{equation}
Here $\lambda$ should be at least the dust scaleheight and also be large enough for diffusion to be ineffective. The last step inserts the scale $\lambda$ where the diffusive mixing time equals the growth timescale, $t_\mathrm{sgi}=\lambda^2/\delta H^2\Omega_K$. The evaluated timescale in \eq{sgi} is rather long for our default parameters; SGI requires relatively massive (high $\Sigma$) or quiescent (low $\delta$) disks.
%In it axisymmetric overdensities|i.e., dust-loaden rings| slowly collapse radially with speeds regulated by the settling velocity.

In contrast, the streaming instability is not gravitational in nature but powered by the background pressure gradient $\eta$. Formally, we can conduct a linear stability analysis on the steady-state flow solutions (like \eq{drift-motions}; \citealt{YoudinGoodman2005}).  The streaming instability can be understood as a subset of more general resonant drag instabilities in which an acoustic sound wave concentrates the dust, which in turn strengthens the perturbing wave \citep{MagnanEtal2024}. Numerical simulations are typically employed to understand when particle-laden fluids exhibit strong clumping. Studies seem to indicate that this threshold lies around $\rho/\rho_\mathrm{gas}\sim1$.

\subsection{Collapse}
Even when clumps form, it is not guaranteed they will collapse.  A key reference is the Roche density
\begin{equation}
    \label{eq:Roche}
    \rho_R \sim \frac{m_\star}{r^3}
    \rightarrow 6\times10^{-7}\ \mathrm{g\,cm^{-3}} %\; (m_\star)_\odot r_\mathrm{au}^{3/2}
\end{equation}
which is the density above which self-gravity is able to overcome disruptive shear motions. Internal dissipation (e.g., inelastic collisions) would then promote further collapse to solid densities. However, in a gas-rich disk $\rho>\rho_R$ may not be a sufficient condition for clumps to stay bound.  More generally, under a two-fluid approximation (where particles and gas are treated separately but linked by drag forces) the clump may yet disperse due to aerodynamic erosion \citep{KlahrSchreiber2021}. Larger clumps are therefore preferred. Once they collapse to solid densities on a free-fall time (${\sim}1/\sqrt{G\rho}$), these clumps need to shed angular momentum, resulting in further fragmentation. A key research line is to compute the size distribution of emerging planetesimal bodies as well as their binarity.

%If a processes self-amplifies, such that particle density increases, it is unstable. Because particles do not feel pressure, solids may be more prone to instability than gas. 

\section{Planetesimals to Planets}
\label{sec:pltsm-planets}

In the next phase of planet formation, solid planetesimal bodies combine into a limited number of planetary bodies by pairwise collisions. Without gravitational focusing, the two-body collision timescale between bodies of size $R_s$ and surface density $\Sigma_s$ is:
\begin{equation}
    \label{eq:tcol}
    t_\mathrm{col-2b} 
    = \left( n_s\sigma_s \Delta v \right)^{-1}
    \sim \frac{R_s\rho_\bullet}{\Sigma\Omega_K}
    = \frac{(R\rho_\bullet)_s r^{3/2}}{\Sigma (Gm_\star)^{1/2}}
    \rightarrow 1.6\times10^3\,\mathrm{yr}.
    %\ \frac{R_\mathrm{km} a_\mathrm{1\,au}^{3/2}\rho_{\bullet} }{\Sigma_{10}  m_{\star,\odot}^{1/2}}.
\end{equation}
Here $\sigma_s=4\pi R_s^2$ is the geometrical cross section, and the bodies are assumed to extend over a height $i\times r$, such that their number density equals $n_s\sim \Sigma_s/irm_s$. I assumed further that the velocity dispersion $\Delta v \sim ev_K$, where $e$ is eccentricity, and energy equipartition ($\Delta v \sim \Delta v_z \sim iv_K$) such that eccentricities and inclinations are similar, $i \sim e$. Finally, I will assume in the following that the small bodies dominate the total solid mass budget, $\Sigma_s\approx\Sigma$.  \Eq{tcol} is also the timescale on which these bodies grow, provided collisions result in sticking (at high $\Delta v$ weak planetesimals are more likely to also fragment). When a large body of size $R_p \gg R_s$ accretes smaller planetesimals its mass-doubling timescale, again in the absence of gravitational focusing, reads:
\begin{equation}
    \label{eq:tgr-geo}
    t_\mathrm{gr,geo} = \frac{m_p}{m_s(n_s\sigma_p \Delta v)} 
    \sim \frac{(\rho_\bullet R)_p}{\Sigma \Omega_K}
    = \frac{(R\rho_\bullet)_p r^{3/2}}{\Sigma (Gm_\star)^{1/2}}
    \rightarrow 8\times10^7\,\mathrm{yr}.
\end{equation}
\Eq{tgr-geo} demonstrates that planets orbiting their stars at distances not much smaller than 1\,au can in fact grow without the aid of gravitational focusing; the dynamical state of the bodies does not matter (note that $\Sigma$ would increase with decreasing $r$). Beyond 1\,au, on the other hand, the growth timescale becomes prohibitively long. To address this issue, \citet{Safronov1972} proposed a bimodel size distribution in which a select number of large bodies|referred to as (planetary) embryos or protoplanets|accrete smaller-sized bodies (pebbles or planetesimals) that make up most of the mass budget, with cross sections $\sigma$ much larger than the geometrical value.  The emergence of a two component system is crucial for the success of planet formation at distances beyond ${\sim}1\,\mathrm{au}$.

Under quiescent conditions, mutual accretion of planetesimals gives rise to a process known as runaway growth, a positive feedback process that results in this desired two-component distribution. Growth nevertheless transitions to the slower oligarchic growth, characterized by negative dynamical feedback. This significantly slows down growth at disk radii beyond $\sim$au. In contrast, accretion of aerodynamically active $\tau_s\sim1$ particles|pebbles (see \se{prelim})| avoids the issue of dynamical feedback, though the emergence of a desired two-component system is less obvious in the case of pebble accretion.

Growth is further limited by isolation masses. For planetesimal accretion, growth is assumed to be local and the protoplanet will eventually consume all material withing an annulus extending several $\sim$Hill radii away from the planet. In pebble accretion protoplanets may open shallow gaps in the disk, precluding pebbles from drifting close enough to the planet. After the gas disk has dispersed, protoplanets may find themselves in closely packed orbits, potentially leading to orbit crossing, where giant impacts restore dynamical stability.

\subsection{Planetesimal Runaway Growth}
\begin{figure}[t]
\centering
\includegraphics[width=\textwidth]{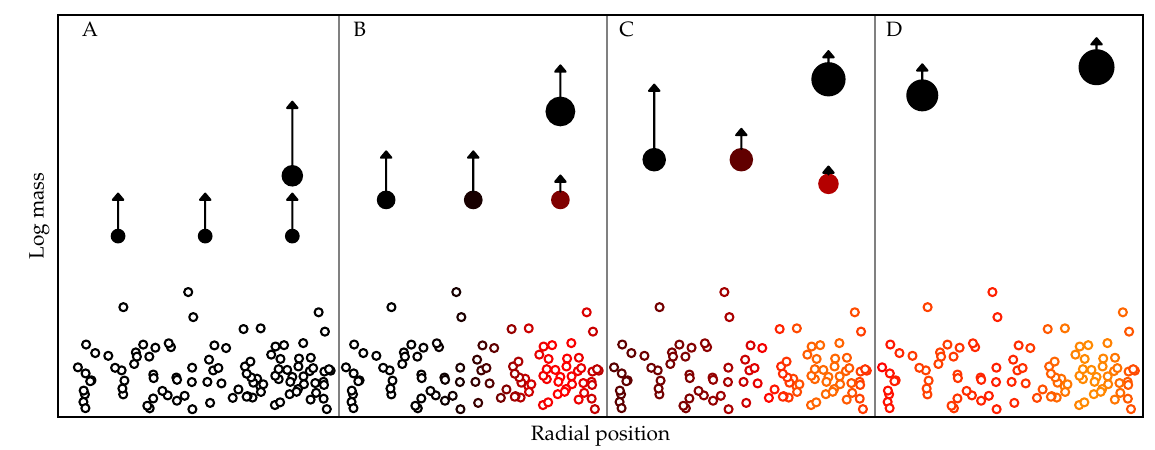}
\caption{\label{fig:oligarchy}Schematic illustrating the transition from runaway to oligarchic growth. The colors represent the dynamical state of the bodies, with lighter shading indicating higher eccentricity. (a) During runaway growth, higher-mass bodies grow at a faster pace than smaller bodies within or between feeding zones. (b and c) Dynamical feedback by the runaway bodies on smaller ones suppresses growth within a feeding zone, allowing large bodies in neighboring zones to catch up. (d) This leads to the emergence of a two-component system (oligarchy), characterized by massive protoplanets and smaller planetesimals.}
\end{figure}

In a two-body encounter, conservation of angular momentum and energy gives the collisional cross section:
\begin{equation}
    \sigma = \sigma_\mathrm{geo} \left( 1 + \Theta \right) 
    = \pi (R_1 + R_2)^2 \left( 1 + \frac{v_\mathrm{esc}^2}{\Delta v^2} \right) 
%   \propto \sigma_\mathrm{geo} \frac{m}{R\Delta v} \qquad \mathrm{if\ } v_\mathrm{esc} \gg \Delta v
\end{equation}
where $v_\mathrm{esc} = \sqrt{2G(m_1+m_2)/(R_1+R_2)}$ is the mutual escape velocity with $\Delta v$ the approach velocity and $\Theta$ the gravitational focusing factor. This expression cannot become infinite, because Keplerian shear effectively puts a floor to the relative velocity: $\Delta v \sim R_H\Omega_K$ and $\sigma \sim RR_H$.  If we for the moment ignore gravitational feedback on the relative velocities, such that $\Delta v$ is constant (i.e., $R_H\Omega_K < \Delta v < v_\mathrm{esc}$, the so-called dispersion-dominated limit), we obtain $\sigma \propto m^{4/3}$ and a growth timescale that decreases with size: $t_\mathrm{gr} \propto m^{-1/3}$. Peculiarly, the pace of growth increases with mass|a runaway process. Additionally, because the mass ration $m_1/m_2>1$ increases with time, $d(m_1/m_2)/dt>0$; more massive bodies outgrow their low-mass competitors. 
%Note that, as I define it, runaway growth is not necessarily fast; it only says that the mass difference diverges.

This phase of decreasing growth timescales ends when the large bodies dynamically excite the planetesimal population.  To understand this, we need to introduce the concept of feeding zone|the region from which a massive body accretes material (see \fg{oligarchy}). Most encounters do not result in collisions. If we define $\sigma_\mathrm{90}$ as the cross section to deflect the relative velocity vector by at least 90$^\circ$, we can find the corresponding cross section, $\sigma_{90} \sim \sigma_\mathrm{geo} (v_\mathrm{esc}/\Delta v)^4$. This process, \textit{viscous stirring}, dynamically heats the bodies in the feeding zone, increasing their rms-values of eccentricities and inclinations. Meanwhile, collisionless encounters also ensure energy equipartition (dynamical friction): bigger bodies will have lower $e$ and $i$ than smaller ones.  

As viscous stirring increasing $\Delta v$, the growth timscale $t_\mathrm{gr}$ of large bodies increases with mass (see below). However, within a single feeding zone, one facet of runaway growth still persist: $d(m_1/m_2)/dt>0$. This means that more massive bodies continue to grow faster than their lower-mass counterparts. As a result, the system evolves into a state where bodies {within the same feeding zones diverge, while masses between different zones converge (\fg{oligarchy}). A two-component system (of big bodies and leftover planetesimals) has been established and becomes more apparent with time. This situation is referred to as \textit{oligarchic growth}. 

The duration of the runaway growth phase|in which growth proceeds at least exponentially with time|lasts on the order of the initial two-body collision timescale among the planetesimal population, \eq{tcol} \citep{OrmelEtal2010}. 
These and other studies typically assume idealized conditions. In reality, the extent of the runaway growth phase may be determined by factors such as the planetesimal mass spectrum and the timescale over which the planetesimals form.

\subsection{Planetesimal Accretion}
Accretion of planetesimals has traditionally been the \textit{modus operandi} for understanding planet formation. However, in the outer disk it can be prohibitively slow.

In the following, typical oligarchic growth assumptions are assumed: a two-component system consisting of protoplanets ``p'', separated by several Hill radii, and smaller planetesimal bodies ``s'' which contain most of the solid mass $\Sigma$. This configuration lends itself to analytical estimates (e.g., \citealt{KokuboIda2000}). Here, I merely highlight key scaling relationships. For example, we can readily write $n{-}\sigma{-}\Delta v$ expressions for the rates of growth, stirring, and gas damping:
\begin{itemize}
    \item \underline{growth} of big bodies due to accreting small planetesimals happens at a rate $t_\mathrm{gr}^{-1} \sim n_s m_s \sigma_\mathrm{col} \Delta v /m_p \sim R_p^2 \Theta \Sigma (\Delta v/ir) /m_p$.
%where the subscript ``s'' refers to the planetesimal population (small bodies) and the ``p'' to the big bodies that are destined to become planets. 
        The dispersion-dominated regime, where approach velocities $\Delta v$ are determined by the eccentricity of the bodies usually applies. Then, $i\sim e$ and $\Delta v/ir \sim \Omega_K$.
    \item \underline{viscous stirring} of the planetesimal population by the protoplanets is due to collisionless encounters: $t_\mathrm{visc}^{-1} \sim n_p \sigma_{90} \Delta v$. 
        %\sim \tilde{b}^{-1} (R^2/R_H a) (v_\mathrm{esc}/\Delta v)^4 (\Delta v/ia)$ 
    The oligarchic occupation rule can be employed to find the number density $n_p$: one single big body takes control of a ``feeding zone'' of width ${\sim}\tilde{b}R_H$ where $\tilde{b}$ is a constant thought to be ${\approx}10$ and stirs all planetesimals over a volume $n_p^{-1} \sim (2\pi r) \times \tilde{b}R_H \times ir$.
    \item \underline{turbulent stirring} is due to density inhomogeneities in the gaseous disk, induced, e.g., by turbulence, which gravitationally torque the planetesimal bodies.% \citep{OkuzumiOrmel2013}\ccc{maybe other ref}. 
    \item \underline{gas drag} will damp the eccentricities and inclinations at a rate $t_\mathrm{gas}^{-1} = C_D \pi R_s^2 \rho_\mathrm{gas}/2m_s \times \Delta v$. In this expression $C_D$ is the gas drag constant, which is order unity if the planetesimals are large.
    \item \underline{collisional damping/fragmentation} among the planetesimals $t_\mathrm{col}$ \eqp{tcol}.
\end{itemize}

For simplicity only proportionality relationships for some parameters are shown in the following. If growth is limited by viscous stirring, equating $t_\mathrm{gr}$ and $t_\mathrm{visc}$ for the relative velocity gives $\Delta v \propto R_p^2 r^{-1} \Sigma^{-1/2}$. This implies that gravitational focussing factors reduce with planet radius, $\Theta \propto R_p^{-2}$ and $t_\mathrm{gr} \propto R_p^3 \Sigma^{-2} r^{-1/2}$. This scaling relationship captivates the problem with planetesimal accretion: the more potent collisionless encounters provide a powerful negative feedback effect, such that growth slows down with $R_p$. For it to succeed, the effective capture radius must be enhanced or the planetesimals' random motions must be damped.  Gas drag alleviates the situation, but only partially. By equating $t_\mathrm{visc}$ with $t_\mathrm{drag}$ we obtain $\Delta v \propto R_p(C_D\rho_\mathrm{gas})^{-1/5} r^{-7/10}$, resulting in gravitational focusing factors independent of radius $R_p$ and growth timescales $t_\mathrm{gr}\propto R_p (C_D \rho_\mathrm{gas})^{-2/5} \Sigma^{-1} r^{1/10}$. Gas damping ameliorates the feedback effect and reduces growth timescales in the inner disk regions. However, it is much less effective in the outer disk, as can be seen when we factor in the implicit dependences of $\rho_\mathrm{gas}$ and $\Sigma$ on $r$. 

A proposed solution to this ``slow growth'' problem would be a more massive disks (which increases both $\Sigma_s$ and $\rho_\mathrm{gas}$) and smaller planetesimals. Smaller planetesimals or fragments of planetesimals offer two advantages: they can be more effectively damped and that they can be captured by the extended proto-atmospheres \citep{InabaIkoma2003}. However, their reduced strength may trigger a collisional cascade \citep{KobayashiEtal2010}, grinding them down to sizes where they become aerodynamically active and causing them to drift out of the planet's feeding zone. The accretion characteristics of these particles share elements of both planetesimal accretion as well as pebble accretion.

Ignoring radial transport, protoplanets will eventually accrete all material in their feeding zone (${\sim}(2\pi r)\tilde{b}R_H \Sigma$) where $\tilde{b}R_H$ is the distance between neighboring protoplanets. The corresponding maximum mass or isolation mass is:
\begin{equation}
    \label{eq:pltsm-iso}
    m_\mathrm{iso}^\mathrm{pltsm} = \frac{(2\pi\tilde{b}\Sigma r^2)^{3/2}}{(3m_\star)^{1/2}}
    \rightarrow 0.11\,m_\oplus%\; a_\mathrm{1\,au}^3 \tilde{b}_{10} \Sigma_\mathrm{10}^{3/2} m_{\star,\odot}^{-1/2}.
\end{equation}
In the inner disk, the low planetesimal isolation masses make giant impacts necessary for further growth.

\subsection{Pebble Accretion}
\label{sec:pebacc}
Pebble accretion describes the aerodynamical-assisted capture of particles in the gravitational well of a protoplanet. These particles must be of the right aerodynamical size: big enough to decouple from gas streamlines, but small enough to dissipate significant amounts of energy to become trapped within the planet's Hill radius. For optimal coupling, $\tau_s \sim 0.1{-}1$, the cross section can approach the planet's Hill radius.  Therefore, pebble accretion rates depend only on the planet mass|not its radius|and pebble accretion lacks close (grazing) encounters.  The self-limitation of growth due to viscous stirring does not exist in pebble accretion; after an encounter gas friction damps the motions of $\tau_s \lesssim 1$ pebbles back to the free drift expressions of \eq{drift-motions}. This lack of dynamical feedback allows pebble accretion to proceed without the slowdown seen in planetesimal accretion.

\begin{figure}[t]
\centering
\includegraphics[width=0.8\textwidth]{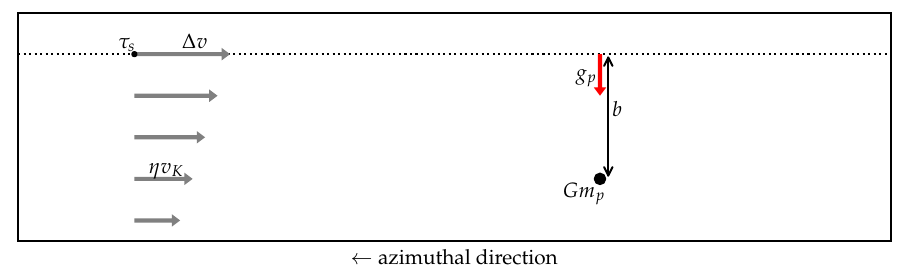}
\caption{\label{fig:pebble-geom}Geometry of the encounter during pebble accretion. Small pebbles approach the planet at a velocity $\Delta v$, determined by the headwind ($\eta v_K$) and, for large impact parameter $b$, the Keplerian shear (${\sim}\Omega_Kb$). At closest approach, the pebble experiences a gravitational acceleration $g_p$ of approximately $Gm_p/b^2$ acting over a timescale $t_\mathrm{enc}\sim b/\Delta v$.}
\end{figure}
The key characteristics of pebble accretion can be understood from the duration of the encounter in relation  to the settling timescale, see \fg{pebble-geom}. The former is determined by the impact parameter $b$ and the velocity $\Delta v$ with which the pebble approaches the planet, $t_\mathrm{enc}\sim b/\Delta v$. The settling time is given by $t_\mathrm{settl} = b/v_\mathrm{settl}$, where the settling velocity follows from force balance ($v_\mathrm{settl}/t_\mathrm{stop}=Gm_p /b^2$ with $t_\mathrm{stop}$ the stopping time).  By equating $t_\mathrm{settl}$ and $t_\mathrm{enc}$ the impact parameter for pebble accretion follows: $b\sim\sqrt{Gm_pt_\mathrm{stop}/\Delta v}$. Yet for pebble accretion to be viable, gas drag should have the opportunity to act during the encounter, $t_\mathrm{stop} \le t_\mathrm{enc}$. This implies that pebble accretion vanishes below a mass ${\sim}t_\mathrm{stop}\Delta v^3/G$. In other words, pebble accretion requires an initiation mass, which, in a finer analysis was determined \citep{VisserOrmel2016}:
\begin{equation}
    \label{eq:Rpa}
    R_\mathrm{pa,init} 
    = 0.31 \Delta v \frac{r^{0.42} \tau_s^{0.28}}{\sqrt{G} \rho_\mathrm{\bullet}^{0.36} m_\star^{0.14}} 
    \rightarrow 140\,\mathrm{km}
%   450\,\mathrm{km}\; (\Delta v)_\mathrm{50\,m/s} \rho_\mathrm{\bullet}^{-0.36} a_\mathrm{au}^{0.42} \tau_s^{0.28}.
\end{equation}
Here $\Delta v$ is the approach velocity between pebbles and planets and $\rho_\bullet$ the internal density of the planetesimal. If the latter is assumed to move on a circular Keplerian orbit $\Delta v^2 = v_r^2 + (v_\phi-v_K)^2 \sim (\eta v_K)^2$ for small bodies and $\tau_s<1$ particles, see \eq{drift-motions}.
Below $R_\mathrm{pa,init}$ there is no aerodynamical capture of pebbles. Instead, pebbles are accreted ballistically, in the same way as planetesimals.  \Eq{Rpa} highlights that pebble accretion requires a sufficiently massive seed, in particular in the outer disk.

When approach velocities are determined by the Keplerian shear, $\Delta v \sim b\Omega_K$, impact parameters for pebble accretion can approach the Hill radius, $b\sim R_H \tau_s^{1/3}$, resulting in cross sections much larger than in planetesimal accretion. In the 2D, shear-dominated limit|the most optimal case|the growth timescale of a planet by pebble accretion becomes:
\begin{equation}
    \label{eq:pa-fastest}
    t_\mathrm{PA-fastest} \sim \frac{m_p}{\left(R_H\tau_s^{1/3}\right)^2\Omega_K\Sigma}
    \sim \frac{m_p^{1/3} m_\star^{1/6}}{\sqrt{G r} \Sigma \tau_s^{2/3}}
    \rightarrow 9\times10^3\,\mathrm{yr}.
\end{equation}
Because $r^{1/2} \Sigma$ is likely only a weak function of disk orbital radius, this expression reveals pebble accretion is particularly potent in the outer disk.  \Eq{pa-fastest} demonstrates the power of pebble accretion, yet it misleads us into thinking that pebble accretion is an efficient process. Paradoxically, the probability ($\epsilon$) of a drifting pebble to be accreted by a planet is low. For example, in this particular limit (2D, shear) the probability of accretion is \citep{LiuOrmel2018}:
\begin{equation}
    \epsilon 
    = \frac{\dot{m}_p}{2\pi r v_r \Sigma}
    = 0.23 \frac{(m_p/m_\star)^{2/3}}{\eta \tau_s^{1/3}} 
    \sim \left( \frac{m_p}{m_\star} \right)^{2/3} h^{-2} \tau_s^{-1/3}
    \rightarrow 0.11
\end{equation}
and $\epsilon$ would be much lower in the outer disk (higher aspect ratio) or when turbulence renders the 2D limit inappropriate.  For each Earth-mass in growth by pebbles, a total pebble mass $m_\oplus/\epsilon$ is required to flow across the planet's orbit. 

Dense particle rings, such as seen with ALMA, provide an environment where the previously mentioned limitations are largely resolved. These rings are manifestations of pressure maxima ($\eta\approx0$) or high midplane densities, which reduce drift velocities ($Z\gg1$ in \eqp{drift-motions}). Under such conditions (i) the planetesimal initiation size $R_\mathrm{pa,init}$ is small due to the reduced $\Delta v$; (ii) pebbles cannot leak away easily; and (iii) $t_\mathrm{gr}$ is short due to the high densities. In addition, if the ring is sufficiently narrow, only one protoplanet can reside within it at any given time (others are scattered away). These conditions support sustained growth at the optimal accretion rate \eqp{pa-fastest}, see \citep{JiangOrmel2023}.

Pebble accretion relies on particles drifting freely from the outer disk to the location of the planet. This drift is interrupted when planets start to open (shallow) gaps in the disk, creating local pressure maxima just exterior to their orbit onto which pebbles converge. As a result, pebble accretion will terminate at the planet gap opening mass, which happens at the mass where the Hill radius exceeds the disk scaleheight $H$:
\begin{equation}
    \label{eq:pebble-iso}
    m_\mathrm{pebble-iso} \sim h^3 m_\star
    \rightarrow 9\,m_\oplus.
\end{equation}
(Turbulence and other effects alter this expression; see \citealt{AtaieeEtal2018} and \citealt{BitschEtal2018i}).

\subsection{Giant Impacts}
A schematic of final planet assembly is given in \fg{assembly}.  The inner disk in particular could end up with a great number of small, Mars-mass planets, due to their low isolation masses (\eqsp{pltsm-iso}{pebble-iso}) and rapid growth times. During the disk phase, efficient tidal damping keeps these bodies on circular orbits. However, once the gas is gone, the gravitational interaction among these bodies would slowly, but surely induce an eccentricity to their orbits. This is a gradual process with an explosive end: orbits will cross and close encounters render the final outcome chaotic. In the outer disk, close encounters would likely result in ejection as the system's escape velocity is low, whereas in the inner disk planets would merge.  The instability time has been numerically determined to exponentially depend on the initial (fractional) separation among the planets with a spacing of 10 mutual Hill radii guaranteeing stability for several billions of years \citep{PetitEtal2020}. Other processes could upset the stability of planet systems sooner, e.g., the sudden disappearance of the primordial disk, perturbations from distant planets or passing stars, an initially high eccentricity, mass loss, etc. An open question in the field is how systems with planets in mean motion resonances|a prediction from planet migration|evolve into an overall non-resonant population that at present day dominate the exoplanet census.

In the very inner disk, the orbital instability timescale may rival that of the disk dispersion. The merger products could then be massive enough to bind the leftover gas. This late gas accretion has been suggested as a way to form gas dwarfs \citep{LeeChiang2016}, see \fg{assembly}.

\section{Gas accretion}
\label{sec:gas}
\begin{figure}[t]
\centering
\includegraphics[width=.8\textwidth]{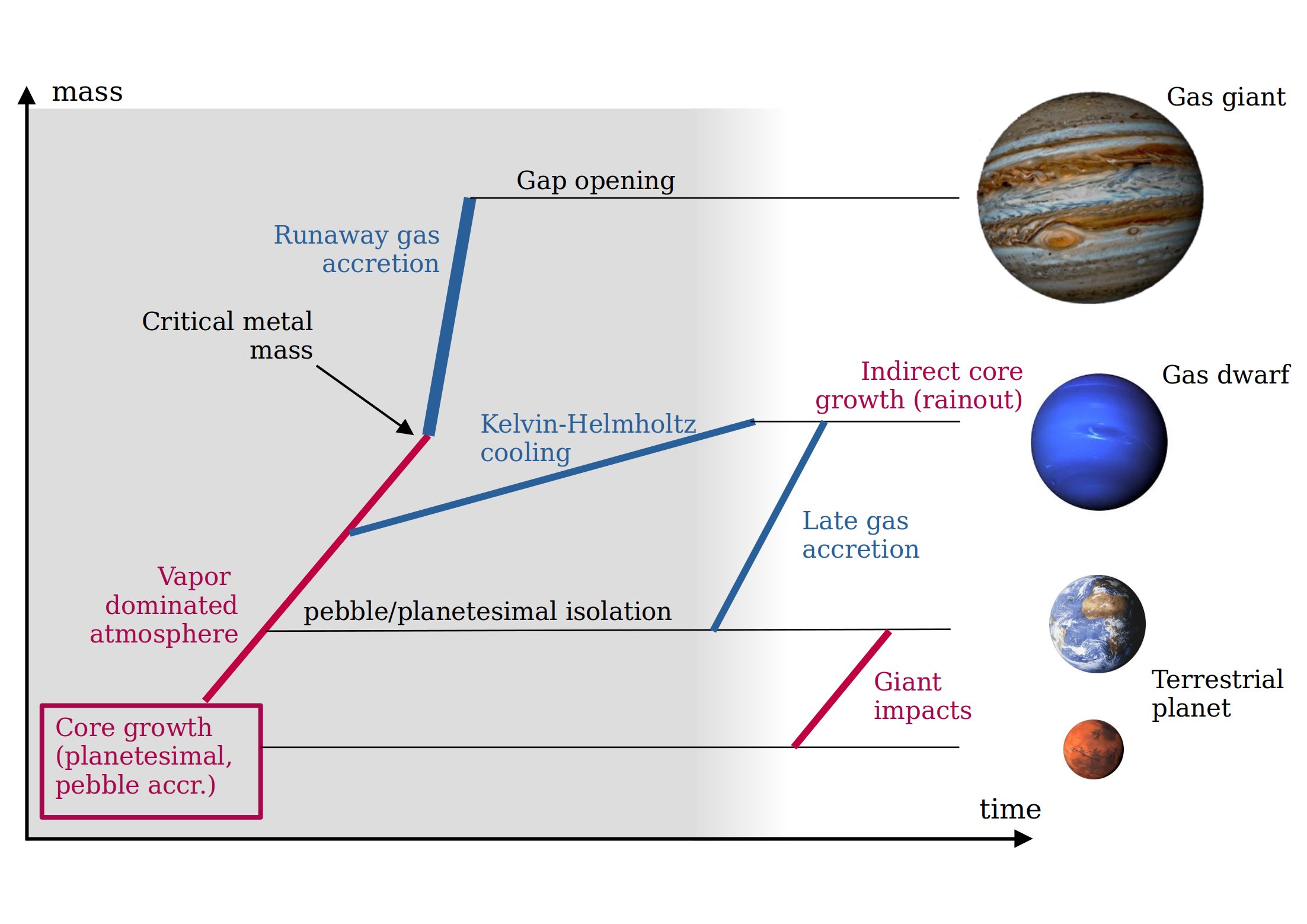}
\caption{Schematic pathways to forming planets in the final stages of the core accretion model. Mechanisms accreting solids are in red, while those involving gas from the disk are in blue.}
\label{fig:assembly}
\end{figure}

In the solar system, planetary atmospheres can be classified as primary and secondary. Primary atmospheres form directly from the hydrogen-helium dominated (H+He) gas mixture of the primordial disk, while secondary atmospheres emerge after the disk dissipates, due to outgassing or volatile delivery. The discussion in this section concerns primary atmospheres. Initially, a planet's atmosphere is continuous with the gas in the disk, as it maintains pressure equilibrium. For simplicity, these atmospheres are referred to as \textit{envelopes}. Compared to the compact atmospheres that planets eventual develop, envelopes are large and tenuous. 

The situation is summarized in \fg{assembly}. Envelopes start to appear at the mass when the escape velocity from the planet surface is similar to the gas thermal velocity, i.e., when a planet's Bondi radius $R_B\equiv Gm_p\overline{m}_\mathrm{gas}/k_BT_\mathrm{disk} = Gm_p/(H\Omega_K)^2$ exceeds its physical radius $R_p$, corresponding to a mass
\begin{equation}
    \label{eq:atmos-start}
    m_\mathrm{env-start} \sim \frac{h^3 m_\star^{3/2}}{\rho_{\bullet,p}^{1/2} r^{3/2}}
    \rightarrow 0.0031\,m_\oplus.
%   m_\mathrm{env-start} \sim 0.01\,m_\oplus\ h_{0.05}^3 (m_\star)_\odot^{3/2} (\rho_\bullet)_3^{-1/2} r_\mathrm{au}^{3/2}.
\end{equation}
As a note in passing, the condition $v_\mathrm{esc}(R) \ge 1v_\mathrm{th}$ marks the onset of envelope growth for planets embedded within a gaseous disk. Once the disk disappears, this criterion becomes more stringent, as molecules in the high-velocity tail of the Maxwellian velocity distribution can escape the planet's gravitational pull more easily. The amount of mass contained in envelopes is determined chiefly by thermodynamical principles. If envelopes can shed entropy, they will contract and accrete more gas from the disk. In addition, envelopes can accrete metals from dissolving solids. Finally, if hydrostatic balance fails, the envelope collapses and planets can accrete gas freely from the disk on dynamical timescales. Giant planets form through this runaway gas accretion process. Eventually, giant planets open gaps in the disk, reducing gas accretion.

\subsection{Envelope growth during ongoing accretion of solids}
Accretional heating (from solids) raises the temperature of the envelope over that of the background disk ($T_\mathrm{disk}$). This means that ices (e.g., \ce{CH4}, \ce{H2O}) can sublimate from the pebbles in the envelope, while the refractory component still makes it to the core. However, as the planet grows larger, even refractories (e.g., silicates) will dissolve entirely in the envelope, halting core growth \citep{BrouwersEtal2018}. For planetesimal accretion, this dissolution may happen at a bit later stage. If accretion of solids continues, the envelope rapidly becomes dominated by dissolved vapor. In fact, under these extreme temperature and pressure conditions, the vapor can reach a supercritical state, blurring the distinction between the ``solid'' core and the ``gaseous'' envelope. Enrichment of the atmosphere in heavy elements adds little pressure support to resist gravitational forces. At a certain point the accumulation of metals (from high-$Z$ elements in the core and the envelope) reaches a critical mass, causing hydrostatic support to fail. The planet then freely accretes gas and can transition to a gas giant.

Most planets are not gas giants. To avoid this outcome, solid accretion could stop, causing growth of the atmosphere to be regulated by its cooling and contraction (see below). Another channel is to return (part) of the material back to the disk through recycling flows. Planets embedded in a disk interact with the disk gas in complex ways. Numerical simulations indicate the existence of recycling flows, where disk material penetrates the envelope (though the extent is debated). These flows can carry vapor from dissolved solids back to the disk. Recycling of the (upper) envelope, therefore, removes volatiles and leaves the planet enriched with refractory material \citep{WangEtal2023}. Recycling of volatiles (ices in, vapor out) provides a pathway to form planets with terrestrial compositions, at or even beyond disk icelines.

Gas dwarfs therefore end up with substantial amounts of dissolved vapor in their envelopes. Over time the envelope may cool and allow the silicate vapor to rain out, providing a second mechanism for the growth of the refractory core \citep{VazanOrmel2023}, see \fg{assembly}.

\subsection{Kelvin-Helmholtz gas accretion}
Accretion of solids can slow down or stop when the planet has reached its pebble or planetesimal isolation mass (\eqsp{pltsm-iso}{pebble-iso}). When the mass exceeds \eq{atmos-start} it becomes possible to accrete significant amounts of H+He gas. If the accreted gas cools efficiently enough to maintain a constant temperature $T$, it is straightforward to show that $\rho_\mathrm{gas}\propto \exp[R_B/r]$. However, gas that enters the envelope adiabatically retains its entropy, increasing $T$, and its density profile follows a power-law: $\rho_\mathrm{gas}\propto r^{1/1-\gamma}$, where $\gamma$ is the heat capacity ratio. In this case the mass of the envelope would be limited. 

To accrete more gas, the planet must cool its envelope by radiating away heat. This process, known as Kelvin-Helmholtz (KH) contraction, is analogous to the pre-main sequence phase of stellar evolution. The rate of KH-contraction is determined by the thermodynamical properties of the atmosphere, chiefly its opacity $\kappa$. A simple estimate of the KH-timescale can be made by assuming constant $\kappa$ and energy flux $L$ throughout the envelope. Under these conditions, and considering radiative energy transport, the mass of the envelope follows from the stellar structure equations:
\begin{equation}
    \label{eq:radiative-zero}
    m_\mathrm{env} \sim \frac{\sigma_\mathrm{SB}}{\kappa L} \left( \frac{Gm \overline{m}_\mathrm{gas}}{k_B} \right)^4
\end{equation}
where $\sigma_\mathrm{SB}$ is Stefan-Boltzmann constant. This is known as the radiative zero solution \citep{Stevenson1982}. Here $m=m_c+m_\mathrm{env}$ with the core mass $m_c$ constant. Inverting this expression, we see that $L$ first decreases with increasing $m_\mathrm{env}$ until the point where $L$ reaches a minimum ($L_\mathrm{min}$) at $m_\mathrm{env}=m/4$, beyond which it increases. The KH-timescale of the envelope is then 
\begin{equation}
    \label{eq:KH}
    t_\mathrm{KH}\equiv \frac{|E|}{L_\mathrm{min}} 
    \sim \frac{Gm_p m_\mathrm{env}}{R_B L_\mathrm{min}}
    = \frac{3k_B^4 T_\mathrm{disk} \kappa}{16\pi \sigma_\mathrm{SB} G^4 \overline{m}_\mathrm{gas} m_p^2}
    \rightarrow 3.4\times10^8\,\mathrm{yr}
    %\sim 10^8\,\mathrm{yr}\;  \kappa_{1} m_\oplus^{-2} (T_\mathrm{disk})_\mathrm{100\,K} \mu_{2.34}^{-5}.
\end{equation}
Numerical calculations that account for convection, non-uniform opacities, and a non-ideal equation of state yield estimates in this range \citep{IkomaEtal2000}.  If $t_\mathrm{KH}$ exceeds the disk lifetime, planets will avoid runaway gas accretion and emerge as gas dwarfs. This is achieved for low $m_p$ or high opacity. However, for sub-Neptune planets ($m\approx10\,m_\oplus$) KH-timescales will become uncomfortably short, because several processes can significantly reduce the opacity|such as the formation of macroscopic bodies from dust, grain growth and sedimentation in the envelopes. Additionally, the presence of dissolved ices and refractories may increase the mean molecular mass $\overline{m}_\mathrm{gas}$, further accelerating contraction. Since there are many more gas dwarfs than gas giants, additional mechanisms may be required to halt KH-contraction. The above-mentioned recycling mechanism, which injects high-entropy gas into the envelope, has been proposed to prevent planets in the inner disk from cooling too fast \citep{OrmelEtal2015i}.

\subsection{Runaway gas accretion}
If hydrostatic support fails, planets can freely accrete from the surrounding disk. For planets embedded in the disk the rate would be limited by the Bondi radius, $\dot{m}_B \sim R_B^2 c_s \Omega_K \rho_\mathrm{gas}$, while planets that have their Hill radius exceeds the disk scaleheight, the rate would be $\dot{m}_\mathrm{2d}\sim R_H^2 \Omega_K \Sigma_\mathrm{gas}$. Converting to a timescale $t_\mathrm{gas}=m/\dot{m}$:
\begin{equation}
    t_\mathrm{gas}^\mathrm{Bondi}
    = \frac{\sqrt{2\pi} h^4 m_\star^{3/2}}{m_p \sqrt{G r} \Sigma_\mathrm{gas}}
    \rightarrow 10^3\,\mathrm{yr}
    ; \qquad
    t_\mathrm{gas}^\mathrm{Hill}
    = \frac{3^{2/3} m_p^{1/3} m_\star^{1/6}}{\sqrt{G r} \Sigma_\mathrm{gas}}
    \rightarrow 42\,\mathrm{yr}
    \label{eq:tgr-BH}
\end{equation}
(the longer timescale applies and we have omitted an intermediate regime, see \citealt{ChoksiEtal2023}.) These expressions indicate gas accretion can occur extremely rapid, with planets consuming all material in their vicinity. In addition, gravitational repulsion of gas further reduces the surface density, making gas-accretion a self-limiting process. 

Gap opening starts when the Hill radius exceeds the pressure scaleheight $H$, which condition is similar to the pebble isolation mass, \eq{pebble-iso}.  The rate at which planets open gaps can be understood from the impulse approximation: gas parcels on distant orbits are repelled by the planet, while turbulent diffusion counteracts gap opening. In general, gap opening, planet accretion, and planet migration occur simultaneously and their intertwined nature make it hard to identify which factors determine the final mass planets end up with. A key factor is whether gas accretes directly to the planet or whether it is funnelled through a circumplanetary disk. If accreted directly, it is likely that material is shock-heated, consistent with the \textit{cold start} scenario usually attributed to core accretion.  Currently, PDS 70 is the only disk that harbors evidence for shock-heated material \citep{HaffertEtal2019}.

%act to determine the accretion rate and final mass of these giant planets. 

\section{Final Thoughts and Outlook}
This review has provided an overview of the mechanisms through which planets can form. However, one key question remains: to which extent do these mechanisms actually materialize during the planet formation era? This is difficult to address as many of these mechanisms are challenging|if not impossible|to observe directly, though some evidence does exist (e.g., grain growth or giant impacts).  An alternative approach to test formation mechanisms is to develop an end-to-end model of planet formation. These models generate synthetic planet populations that can be compared with the observed exoplanet census. Such population synthesis models should include factors not covered here: dynamical interactions with the gas (planet migration) and interactions among planets (e.g., resonance trapping). In addition, it also necessitates a solid understanding of how protoplanetary disks form as a byproduct of star formation and how they evolve over time. The appeal of the population synthesis approach is its potential to identify processes that generally govern planet formation. However, the downside is that the accuracy of the entire modelling chain is highly dependent on the weakest link in our understanding of planet formation and disk evolution.

\medskip
\noindent
I end this chapter with outlining several open questions in planet formation and how they are connected to planet formation mechanims:
\begin{itemize}
    \item Is disk instability (DI) necessary to explain the presence of directly imaged planets? A persistent problem in the core accretion scenario is to explain the existence of planets beyond ${\approx}10\, \mathrm{au}$. In particular, directly imaged planets are found to orbit stars at several tens of au, where processes involved in standard planetesimal accretion, such as oligarchic growth, appear too slow for planet formation. However, pebble accretion in rings offers an alternative for forming planet cores at these distances. If pebble accretion is this effective, is there still need for DI?
    \item Is planet(esimal) formation a disk-wide phenomenon or does it favor specific locations?  One of the key bottlenecks to growth is the formation of planetesimals, because at sizes below planetesimals there is no obvious sticking agent|both surface forces and self-gravity are too weak.  If planets do form throughout the disk, it may indicate planetesimals form through direct growth, which does not require special conditions, or through a widespread instability. Conversely, if planetesimal formation requires special conditions|such as an elevated solids-to-gas ratio|this would imply that planetesimals, and thus planets, form only at specific locations, e.g., icelines.
    \item What determines the final (gas) mass of planets? Is the gas mass in planets limited by thermodynamical principles (e.g., the Kelvin-Helmholtz timescale), hydrodynamical principles (e.g., recycling), or do planets simply accrete all the gas in their vicinity? These processes may depend on environment factors, e.g., disk radius.
    \item Do planet systems exhibit different classes of architectures? While over 5,000 exoplanets have offered some insights, it remains unclear whether planetary systems can be categorized into specific architectures (e.g., solar-system like or similar-mass planets). If distinct classes exist, an intriguing question is whether they can be linked to specific planet formation mechanisms.
\end{itemize}

\begin{ack}[Acknowledgments]
    \
    I acknowledge support by the National Natural Science Foundation of China (grants no. 12250610189 and 12233004).  I would like to thank my group members at Tsinghua University for proof-reading the text.
\end{ack}

\seealso{article title article title}

\bibliographystyle{Harvard}
\bibliography{reference,ads,arXiv}

\end{document}